\documentclass[conference]{IEEEtran}
\IEEEoverridecommandlockouts
\usepackage{amsmath,amsfonts}
\usepackage{algorithmic}
\usepackage{algorithm}
\usepackage{array}
\usepackage[caption=false,font=normalsize,labelfont=sf,textfont=sf]{subfig}
\usepackage{textcomp}
\usepackage{stfloats}
\usepackage{url}
\usepackage{verbatim}
\usepackage{graphicx}
\usepackage{cite}
\usepackage{caption}

\begin{document}

\title{Mobility-Aware Power Control for VCSEL-Based Indoor OWC\\
}

\author{
    Walter Zibusiso Ncube,~\IEEEmembership{Student Member,~IEEE},
    Ahmad Adnan Qidan,~\IEEEmembership{Member,~IEEE}, \\
    Taisir El-Gorashi,~\IEEEmembership{Member,~IEEE},
    and Jaafar M. H. Elmirghani,~\IEEEmembership{Fellow,~IEEE}
    \thanks{Walter Zibusiso Ncube,
    Ahmad Adnan Qidan, Taisir El-Gorashi,
    and Jaafar M. H. Elmirghani, Department of Engineering, King’s College London, London, United Kingdom (emails: (walter.ncube, ahmad.qidan, taisir.elgorashi, jaafar.elmirghani)@kcl.ac.uk).}
}

% The paper headers
%\markboth{Journal of \LaTeX\ Class Files,~Vol.~14, No.~8, August~2021}%
%{Shell \MakeLowercase{\textit{et al.}}: A Sample Article Using IEEEtran.cls for IEEE Journals}

%\IEEEpubid{0000--0000/00\$00.00~\copyright~2021 IEEE}
% Remember, if you use this you must call \IEEEpubidadjcol in the second
% column for its text to clear the IEEEpubid mark.

\maketitle

\begin{abstract}
Optical wireless communication (OWC) is a promising technology for supporting data-intensive services in indoor environments due to its large unregulated spectrum, high spatial reuse, and potential for multi-gigabit data rates. In particular, vertical-cavity surface-emitting laser (VCSEL) based systems enable highly directional transmission, allowing efficient spatial separation of users and improved link performance. However, the use of narrow optical beams also makes system performance highly sensitive to user mobility and device orientation, as movement directly affects beam alignment and optical channel gain. Consequently, power allocation strategies that ignore mobility dynamics often provision excess optical power to maintain reliable connectivity, resulting in inefficient energy use. In this work, a power control framework for dynamic indoor OWC networks that explicitly accounts for mobility-driven channel variation is developed. It uses a hybrid Gauss–Markov and learning-based approach that captures both user movement continuity and behaviour-driven orientation changes. The mobility states are then used to guide power allocation decisions. Simulation results show that incorporating mobility-aware channel prediction enables more accurate power allocation, and improves energy efficiency compared with conventional power control schemes in dynamic indoor environments.
 
\end{abstract}

\begin{IEEEkeywords}
Optical wireless communication, VCSELs, energy-efficiency, mobility modelling, machine learning.
\end{IEEEkeywords}

\section{Introduction}

\IEEEPARstart{O}{ptical} wireless communication (OWC) has emerged as a promising complement to traditional radio-frequency (RF) systems for meeting the growing demand for high-capacity wireless connectivity. Applications such as immersive multimedia, augmented and virtual reality, and dense Internet of Things (IoT) deployments require multi-gigabit data rates and low latency. OWC addresses these requirements in indoor environments by leveraging the vast unregulated optical spectrum, inherent security due to signal confinement, and directional transmission that enables high spatial reuse \cite{Weng,Walt1}. Recent advances in optical transmitters have further enhanced OWC capabilities. In particular, vertical-cavity surface-emitting lasers (VCSELs) offer high modulation bandwidth, energy efficiency, and scalability through array integration. Unlike LED-based visible light communication systems, VCSELs generate highly directional narrow beams with near-Gaussian profiles, enabling precise user separation and increased link capacity \cite{sarbazi2020tbsystem,Walt2}. VCSEL arrays support parallel beam transmission, achieving multi-gigabit-per-second per link and aggregate terabit-per-second capacities in indoor environments \cite{WaltAhrar,VCSEL}. However, the strong directionality of the VCSEL beams also introduces significant challenges for network control and resource allocation.

In OWC systems, the received signal strength depends strongly on the receivers’s position relative to the transmitter \cite{8540452,9145333}. As a result, user mobility and device orientation variations directly affect the optical channel gain and link reliability. Even small changes in user position or device tilt can lead to significant degradation in received power due to beam misalignment. This sensitivity makes mobility a critical factor in the design of OWC systems, particularly for VCSEL-based networks where narrow beams amplify alignment constraints.

Mobility modelling has been widely studied in wireless networks, with common models including Random Waypoint (RWP) and Gauss–Markov (GM). The RWP model represents movement as randomly selected destinations and speeds \cite{11380140}, while the GM model introduces temporal correlation in velocity and direction, producing smoother and more realistic trajectories \cite{RWP_GM}. The GM model has also been applied to mobile-to-mobile channel modelling to capture non-stationary behaviour due to time-varying motion \cite{GM_channel}. In \cite{SMP}, Semi-Markov models were used to predict user location and dwell time, enabling accurate proactive resource allocation. However, stochastic models such as GM assume Gaussian variations and cannot capture structured behavioural patterns, including abrupt turns, pauses, and device orientation changes, which are critical in OWC systems.
Recent advances in machine learning have enabled data-driven mobility prediction. In \cite{ANN_mob}, a long short-term memory (LSTM) network was used to learn temporal movement patterns more effectively than conventional neural networks. In \cite{ANN_mob1}, Q-learning was applied to model indoor mobility, achieving higher prediction accuracy by adapting to environmental changes. Furthermore, \cite{ANN_mob2} proposed a federated learning framework that incorporates mobility into network optimisation, improving performance and adaptability. However, purely learning-based approaches, while accurate, require large training datasets and may lack robustness under dynamic or unseen conditions. In \cite{GM-LSTM}, we developed a hybrid mobility model that combines a stochastic model with a learning based model to capture temporal and non-linear
mobility dynamics. The hybrid model achieved improved prediction accuracy and performance compared to traditional RWP and GM models.

Resource allocation is a key aspect of OWC networks. A substantial body of work has investigated resource allocation and optimisation in OWC networks, including convex optimisation techniques, heuristic algorithms, and machine learning-based approaches for power control and user association \cite{QL,Ahrar}. In \cite{ANN4}, a model-free deep learning approach was proposed, where a neural network learns optimal allocation policies and achieves near-optimal performance without explicit channel modelling or optimisation solvers. In \cite{ANN2}, centralised optimisation was used to jointly allocate access points and IRS elements, maximising the system sum rate under blockage conditions. Furthermore, \cite{ANN1} applied artificial neural networks to learn resource allocation decisions, improving rate performance compared to conventional methods.
However, most existing approaches assume static or quasi-static user locations and perfect channel knowledge, and therefore do not capture mobility-induced channel variations. %In addition, existing work has not explicitly integrated hybrid mobility prediction models, that combine stochastic and learning-based approaches, into resource allocation frameworks for OWC systems.

To address these limitations, unlike previous work, this paper proposes a mobility-aware power control (MAPC) framework for indoor VCSEL-based OWC systems. The framework is built on a hybrid mobility model that combines the GM process, which enforces temporally correlated and physically realistic motion dynamics, with data-driven learning to capture behaviour-driven dynamics. An LSTM network captures non-linear mobility patterns such as abrupt changes in direction and device orientation, improving prediction accuracy. The predicted mobility states are then used to estimate future channel conditions. These estimates are then provided to a convolutional neural network (CNN) for power allocation. The CNN is trained using optimal solutions from a mixed-integer linear programming (MILP) formulation to solve the energy efficiency (EE) optimisation problem in real time. This enables fast, near-optimal decisions. Results show that the proposed framework improves energy efficiency while maintaining reliable performance under dynamic user mobility.

To the best of the authors' knowledge, this is the first work to jointly integrate a hybrid Gauss Markov-Artificial Neural Network (GM-ANN) mobility model with power control in VCSEL-based OWC systems. This addresses a critical gap in mobility-aware resource allocation for next-generation indoor optical networks. The rest of this paper is organised as follows: Section \ref{System Model} gives the proposed system model. Section \ref{Hybrid Gauss--Markov Framework} describes the proposed hybrid GM-ANN solution and Section \ref{Optimisation Problem} describes the proposed mobility aware power control. The performance analysis of the proposed system is given in Section \ref{Performance Evaluation}, and the conclusions are presented in Section \ref{Conclusion and Discussion}.

\section{System Model}
\label{System Model}
This section presents the system model and the proposed mobility framework. We first describe the indoor OWC network and channel model. Next, we define the user mobility model, including the proposed hybrid GM-ANN approach.
\subsection{Room and System Setup}

Consider an indoor laser-based OWC network designed to provide high data rate communication. A set of optical access points (APs), $\mathcal{A},  a = \{1, \dots, A\}$, are mounted on the ceiling. Each AP is a $C \times C$ micro-lens VCSEL array for optical transmission. The APs provide coverage for a set of users, $\mathcal{U}, u = \{1, \dots, U\}$, located on the communication plane. The users are distributed randomly and vary over time due to mobility. Each user is equipped with an angle diversity receiver (ADR). Users may arrive, depart, and move within the coverage region over time and multi-user interference is mitigated using zero-forcing (ZF) precoding. All APs are connected via WiFi uplink to a central unit (CU) that performs network coordination, mobility estimation, and power allocation. It should be noted that unlike LED-based VLC systems, VCSEL transmitters produce highly directional beams with approximately Gaussian intensity profiles. Consequently, the optical channel gain depends strongly on the spatial relationship between the transmitter and receiver, including propagation distance and beam alignment \cite{liu,RWP}. Small variations in user position or device orientation can therefore lead to significant changes in received optical power.
%\begin{figure}[h]
%\centering 
%\includegraphics[width=1\columnwidth]{Indoor OWC.png}
%\caption{Indoor OWC Environment} 
%\label{fig:Indoor_OWC} 
%\end{figure}
In this work, time is divided into discrete intervals of duration $\Delta t$. At each interval, the system updates the user mobility state, estimates the channel conditions, and computes the transmit power for the next time slot. %This discrete-time model enables tractable optimisation while capturing the dynamic behaviour of the system.

%The achievable data rate for user $u$ served by AP $a$ at time $t$ is given by
%\begin{equation}
%R_{u,a}(t) = B \log_2 \left(1 + \frac{P_{u,a}(t) H_{u,a}(t)}{N_0}\right),
%\end{equation}
%where $B$ is the system bandwidth, $P_{u,a}(t)$ is the transmit power allocated to user $u$, $H_{u,a}(t)$ is the optical channel gain, and $N_0$ is the noise power.

\subsection{Mobility}
%In the considered indoor OWC system, as users traverse the communication floor, they may transition between the coverage regions of different access points, leading to variations in link quality due to changes in relative geometry and alignment. These variations are highly sensitive to both user position and device orientation, as optical links rely on directional transmission and reception. 
\begin{figure}[t]
\centering 
\includegraphics[width=1\columnwidth]{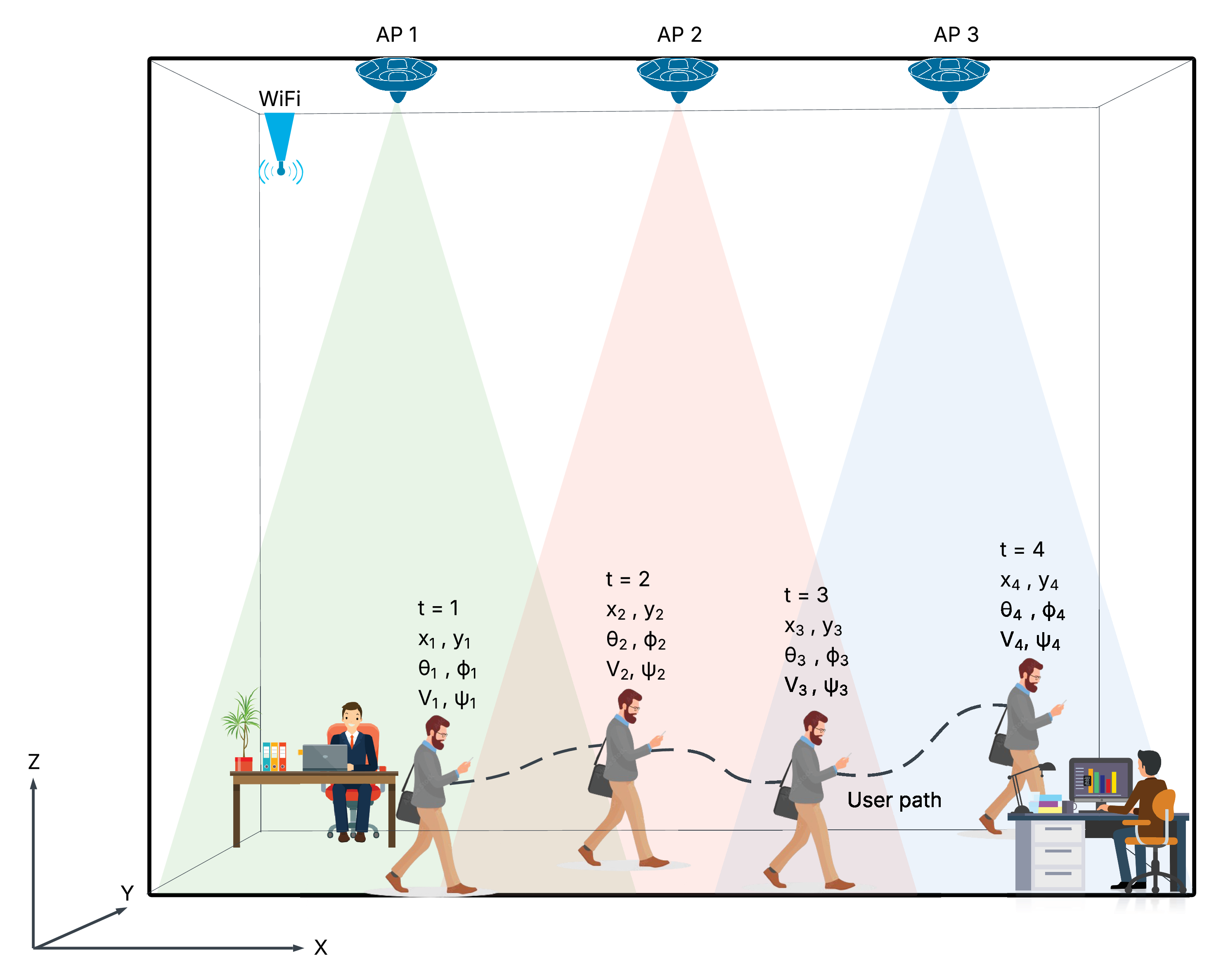}
\caption{User Mobility.} 
\label{fig:Indoor_Mobility} 
\end{figure}
Consider user $u$, as this user moves, as shown in Fig. \ref{fig:Indoor_Mobility}, several parameters evolve over time. The spatial location of the user changes continuously, affecting the propagation distance and line-of-sight conditions. At the same time, the user’s speed and direction determine the trajectory of motion, which governs how rapidly the channel conditions vary. In addition, the orientation of the user device changes and follows
Laplace distributions when users are sitting and Gaussian distributions when walking \cite{RWP}. This introduces fluctuations in the receiver’s field-of-view alignment with the incoming optical beam. These combined effects result in dynamic variations of the optical channel gain. To capture these dynamics, mobility is defined as the joint temporal evolution of user position, motion, and device orientation. Accordingly, the mobility state of user $u$ at time $t$ is expressed as
\begin{equation}
\mathbf{s}_u(t) = [x_u(t), y_u(t), v_u(t), \psi_u(t), \theta_u(t), \phi_u(t)]^T,
\end{equation}
where $(x_u(t), y_u(t))$ denotes the user position, $v_u(t)$ is the instantaneous speed, and $\psi_u(t)$ represents the direction of motion. The device orientation is characterised by the elevation angle $\theta_u(t)$, which defines the tilt of the receiver with respect to the vertical axis and determines the incidence angle of the received optical signal, and the azimuth angle $\phi_u(t)$, which represents the rotation of the device around the vertical axis.

This state representation forms the basis for modelling user mobility in the system and serves as the foundation upon which the proposed hybrid GM-ANN model is developed.

\subsection{Optical Channel Gain}

The optical intensity distribution of a single VCSEL beam at a distance $z$ is given by
\begin{equation}
I(r,z) = \frac{2 P_{\text{out}}}{\pi w^2(z)} \exp\left(-\frac{2r^2}{w^2(z)}\right),
\end{equation}
where $P_{\text{out}}$ is the transmitted optical power, $r$ is the radial distance from the beam centre, and $w(z)$ is the beam radius at distance $z$. The micro-lens modifies the beam divergence according to the ABCD transformation \cite{ABCD}, while preserving the Gaussian shape. The received optical power from a single VCSEL within a receiver aperture of radius $r_0$ is:
\begin{equation}
P_v = P_{\text{out}} \left(1 - \exp\left(-\frac{2 r_0^2}{w_l^2(z)}\right)\right),
\end{equation}
where $w_l(z)$ is the beam radius after the micro-lens transformation. The total transmit power from AP $a$ is
\begin{equation}
P_a = \sum_{i=1}^{c \times c} P_{i,v},
\end{equation}
where $P_{i,v}$ is the transmitted power from the $i$-th VCSEL element.
The optical channel gain between AP $a$ and user $u$ at time $t$ is given by

\begin{equation}
\label{channel_gain}
H_{u,a}(t) =
\begin{cases}
\frac{A_r}{d_{u,a}^2(t)} \Gamma_{u,a}(t)\cos(\theta_u(t)),
& \theta_u(t) \leq \theta_{\text{FOV}} \\
0, & \text{otherwise}
\end{cases}
\end{equation}

\noindent where $\Gamma_{u,a}(t) = \exp\!\left(-\frac{2 r_{u,a}^2(t)}{w^2(d_{u,a}(t))}\right)$, $A_r$ is the receiver area, $d_{u,a}(t)$ is the distance between AP $a$ and user $u$, $r_{u,a}(t)$ is the radial displacement from the beam axis, $w(d)$ is the beam radius at distance $d$, $\theta_u(t)$ is the incidence angle determined by device orientation, and $\theta_{\text{FOV}}$ is the receiver field-of-view.

\section{Proposed Methodology for Dynamic User Behaviour}
\label{Hybrid Gauss--Markov Framework}
In our previous work in \cite{GM-LSTM}, a mobility model was developed that jointly captured temporal correlation and non-linear mobility behaviour in dynamic indoor OWC environments, demonstrating improved performance over traditional RWP and GM schemes. The model combines stochastic and learning-based approaches in a GM-ANN framework. It combines the temporal correlation properties of the GM model with the learning capability of a neural network to capture both stochastic motion dynamics and structured behavioural patterns. Specifically, the GM component models the continuous evolution of user motion, while the learning-based component refines the mobility representation by accounting for non-linearities and user-specific variations that cannot be captured by purely stochastic models as in \cite{GM-LSTM}. Based on the GM model, the velocity and direction evolve as
\begin{align}
v_u(t+1) &= \alpha v_u(t) + (1-\alpha)\bar{v} + \sqrt{1-\alpha^2} w_v(t), \\
\psi_u(t+1) &= \alpha \psi_u(t) + (1-\alpha)\bar{\psi} + \sqrt{1-\alpha^2} w_\psi(t),
\end{align}
where $\alpha \in [0,1]$ is the memory parameter, $\bar{v}$ and $\bar{\psi}$ are mean velocity and direction, and $w_v(t)$ and $w_\psi(t)$ are Gaussian noise terms. The position is updated as: 
\begin{align}
x_u(t+1) &= x_u(t) + v_u(t+1)\cos(\psi_u(t+1))\Delta t, \\
y_u(t+1) &= y_u(t) + v_u(t+1)\sin(\psi_u(t+1))\Delta t.
\end{align}

The orientation angles evolve over time as:

\begin{align}
\theta_u(t+1) &= \alpha_\theta \theta_u(t) + (1-\alpha_\theta)\bar{\theta} + w_\theta(t), \\
\phi_u(t+1) &= \alpha_\phi \phi_u(t) + (1-\alpha_\phi)\bar{\phi} + w_\phi(t),
\end{align}
where $\alpha_\theta, \alpha_\phi \in [0,1]$ are memory parameters, $\bar{\theta}$ and $\bar{\phi}$ are the mean orientation angles, and $w_\theta(t) \sim \mathcal{N}(0,\sigma_\theta^2)$ and $w_\phi(t) \sim \mathcal{N}(0,\sigma_\phi^2)$ are zero-mean Gaussian noise terms. The predicted elevation angle $\theta_u(t+1)$ directly affects the optical channel gain through the incidence angle term, while $\phi_u(t+1)$ influences the effective reception through the ADR.

%To capture behaviour-driven variations such as sudden device tilting, a neural network is used to correct the Gauss--Markov prediction. The final orientation is given by
%\begin{align}
%\theta_u(t+1) &= \theta_u^{\mathrm{GM}}(t+1) + \Delta \theta_u(t+1), \\
%\phi_u(t+1) &= \phi_u^{\mathrm{GM}}(t+1) + \Delta \phi_u(t+1),
%\end{align}
%where $\Delta \theta_u(t+1)$ and $\Delta \phi_u(t+1)$ are the correction terms predicted by the neural network.

A neural network is used to correct the prediction error of the GM model. In this work, the correction network is implemented using an LSTM architecture, which is well suited for modelling sequential data. Unlike feedforward networks, the LSTM can capture temporal dependencies in user mobility by learning from a sequence of past mobility states. This is important because human movement and device orientation changes are inherently time-correlated and often depend on recent motion patterns.

The input to the LSTM consists of a sequence of previous mobility states, including user position, velocity, direction, and device orientation. Based on this sequence, the LSTM outputs a correction term $\Delta \mathbf{s}_u(t+1)$, which represents the difference between the GM prediction and the actual user movement. This allows the model to capture behaviour-driven dynamics such as sudden turns, pauses, and device tilting that are not represented by the GM model. The LSTM is trained offline using synthetic mobility data. The dataset is generated by combining GM trajectories with additional behaviour-driven variations, including abrupt direction changes, speed fluctuations, and orientation dynamics, to emulate realistic indoor user movement. The training objective is to minimise the error between the corrected prediction and the true mobility state. Once trained, the LSTM is used during runtime to refine the GM prediction and improve the accuracy of the predicted mobility state. The final predicted mobility state is given by
\begin{equation}
\mathbf{s}_u(t+1) = f_{\mathrm{GM}}(\mathbf{s}_u(t)) + \Delta \mathbf{s}_u(t+1).
\end{equation}

%The predicted position $(\hat{x}_u(t+1), \hat{y}_u(t+1))$ is used to compute $d_{u,a}(t+1)$, the distance between AP $a$ and user $u$, and $\hat{r}_{u,a}(t+1)$, the radial displacement from the beam axis, while the predicted orientation $\hat{\theta}_u(t+1)$ determines the incidence angle. 

User association can now be determined based on predicted user locations and device orientations. First, the channel gain for each AP–user pair is computed, while links violating the receiver FoV constraint are discarded. Then, an association policy is applied whereby each user is assigned to only one AP that provides the maximum channel gain, and  $\mathcal{U}_a, u = \{1, \dots, U_{a}\}$, denotes the users associated with AP $a$. This scheme effectively captures the impact of spatial configuration and orientation on link quality, providing an effective basis for the subsequent power allocation stage aimed at improving EE.

%Therefore, the channel gain $H_{u,a}(t+1)$ is directly determined by the predicted mobility state. These predicted channel conditions are then used as inputs to a CNN-based power allocation algorithm. The CNN is trained on MILP optimal solutions to enable proactive adaptation of transmit power to anticipated mobility-induced variations.

%The distance $d_{u,a}(t)$ and radial displacement $r_{u,a}(t)$ at time $t$ are given by:
%\begin{equation}
%d_{u,a}(t) = \sqrt{(x_u(t)-x_a)^2 + (y_u(t)-y_a)^2 + (z_u - z_a)^2}
%\end{equation}

%\noindent and,

%\begin{equation}
%r_{u,a}(t) = \sqrt{(x_u(t)-x_a)^2 + (y_u(t)-y_a)^2}
%\end{equation}

%\noindent respectively. Where $(x_u(t), y_u(t), z_u)$ is the position of user $u$ and $(x_a, y_a, z_a)$ is the position of AP $a$.

%\subsection{Achievable Data Rate}

To determine the achievable data rate of each user, let the hybrid GM-ANN model output the predicted user state for the next time slot as:

\begin{equation}
\label{estimated_values}
\hat{\mathbf{s}}_u(t+1)=
\left[
\begin{array}{c}
\hat{x}_u(t+1),\ \hat{y}_u(t+1),\ \hat{v}_u(t+1) \\
\hat{\psi}_u(t+1),\ \hat{\theta}_u(t+1),\ \hat{\phi}_u(t+1)
\end{array}
\right]^T
\end{equation}

where $\hat{x}_u(t+1)$ and $\hat{y}_u(t+1)$ are the predicted user coordinates, $\hat{v}_u(t+1)$ is the predicted speed, $\hat{\psi}_u(t+1)$ is the predicted direction of motion, and $\hat{\theta}_u(t+1)$ and $\hat{\phi}_u(t+1)$ are the predicted device orientation angles. Using the predicted position and orientation, the predicted link distance and predicted radial displacement are given by

\begin{equation}
\label{estimated_d}
\begin{aligned}
& \hat{d}_{u,a}(t+1)=  \\ 
&\quad \sqrt{
\left(\hat{x}_u(t+1)-x_a\right)^2
 + \left(\hat{y}_u(t+1)-y_a\right)^2 + \left(z_u-z_a\right)^2
}
\end{aligned}
\end{equation}

\noindent and
\begin{equation}
\label{estimated_r}
\hat{r}_{u,a}(t+1)=\sqrt{\left(\hat{x}_u(t+1)-x_a\right)^2+\left(\hat{y}_u(t+1)-y_a\right)^2},
\end{equation}
respectively, where $(x_a,y_a,z_a)$ is the position of AP $a$ and $(\hat{x}_u(t+1),\hat{y}_u(t+1),z_u)$ is the predicted position of user $u$.

%\begin{equation}
%\begin{aligned}
%& \hat{H}_{u,a}(t+1)= \\
%&\quad \begin{cases}
%\displaystyle
%\frac{A_r}{\hat{d}_{u,a}^2(t+1)} 
%\Gamma_{u,a}(t+1)
%\cos\!\big(\hat{\theta}_u(t+1)\big),
%& \hat{\theta}_u(t+1)\leq \theta_{\mathrm{FOV}} \\
%0, & \text{otherwise}
%\end{cases}
%\end{aligned}
%\end{equation}

%\noindent where $\Gamma_{u,a}(t+1) =
%\exp\!\left(
%-\frac{2\hat{r}_{u,a}^2(t+1)}
%{w^2\!\big(\hat{d}_{u,a}(t+1)\big)}
%\right).$
Based on equations (\ref{estimated_values}) to (\ref{estimated_r}), equation (\ref{channel_gain}) is updated and the predicted optical channel gain between AP $a$ and user $u$ at time $t+1$ is then expressed as $\hat{H}$. The resulting predicted achievable data rate for user $u$ at time $t+1$ is given by
\begin{equation}
\hat{R}_{u}(t+1)=
B\log_2\left(
1+\frac{(P_{u}(t+1)R_{PD}\hat{H}_{u,a}(t+1))^2}{\sigma^2 +\sum_{\substack{a' \in A \\ a' \neq a}} I_{a'}} 
\right),
\end{equation}

where $B$ is the system bandwidth, $P_{u}(t+1)$ is the  power allocated to user $u$ by AP $a$ for the next time slot, $R_{PD}$ is the receiver responsivity, $\sigma^2$ is the receiver noise current variance, and $\sum_{\substack{a' \in A \\ a' \neq a}} I_{a'}$ denotes the total inter-AP interference, where   $I_{a'}= \sum_{\substack{u \in \mathcal{U}_{a'}}}   (R_{\text{PD}}\hat{H}_{u,a'}P_{u})^{2}$ is the interference from AP $a'$.

\section{Mobility-Aware Power Control}
\label{Optimisation Problem}
This section formulates the power allocation optimisation problem using predicted user states and channel conditions. At each time slot $t$, the system predicts the mobility state and channel gain for each user for time $t+1$. The transmit power is then optimised for the next time slot to maximise the system $EE$ defined as the ratio of the total achievable data rate to the total transmit power consumption: 

\begin{equation}
EE =  \frac{\sum_{u \in \mathcal{U}}  \log\big(\hat{R}_{u}(t+1)\big)}{\sum_{u \in \mathcal{U}} P_{u}(t+1)}
\end{equation}

The objective is to maximise the $EE$ as:

\begin{equation}
\max_{P_{u}}EE
\end{equation}

subject to

\vspace{-0.5em}
\begin{align}
%\sum_{a \in \mathcal{A}} S_{u,a}(t+1) &\leq 1, && \forall u \tag{C1} \label{C1}\\
\hat{R}_{u}(t+1) &\geq R_u^{\min}, && \forall u \tag{C1} \label{C1}\\
\sum_{u \in \mathcal{U}_{a}} P_{u}(t+1) &\leq P_a, && \forall a \tag{C2} \label{C2}\\
P_{\min} \leq P_{u}(t+1) &\leq P_{\max} \tag{C3} \label{C3} \\
P_{\min} &\geq 0 \tag{C4} \label{C4}
\end{align}

\text
(\,\ref{C1}\,) guarantees quality-of-service (QoS) for each user, where $R_u^{\min}$ denotes the minimum required data rate. (\,\ref{C2}\,) ensures that the sum of the power allocated to all users served by AP $a$ does not exceed the AP capacity $P_a$. (\,\ref{C3}\,) ensures that the transmit power allocated to each user lies within a valid range, bounded by $P_{\min}$ and $P_{\max}$. $P_{\max}$ is determined by safety and hardware constraints, while the lower bound ensures reliable signal detection. Finally, (\,\ref{C4}\,) enforces the non-negativity of the power allocation variables.

The formulated optimisation problem is non-convex. Its computational complexity grows exponentially with the number of users. As a result, real-time optimisation becomes intractable. To address this, a learning-based approach is adopted in which a CNN is trained to approximate the optimal solution. The CNN takes as input the predicted channel gain matrix obtained from the hybrid GM–ANN model, along with the user demands and the AP capacities. It then outputs the corresponding transmit power allocation by approximating the solution to the EE optimisation problem. Training data is generated offline by solving the optimisation problem using a MILP formulation under diverse network conditions, providing labelled input-output pairs for supervised learning. The CNN consists of three convolutional layers with 32, 64, and 128 filters, each using $3 \times 3$ kernels and ReLU activation, followed by $2 \times 2$ max pooling. The extracted features are passed through a fully connected layer with 128 units and a linear output layer for power allocation. The network is trained using the Adam optimiser with a learning rate of $10^{-3}$ and mean squared error (MSE) loss over 100 epochs with a batch size of 64. During inference, the trained CNN directly maps predicted channel states to power allocation decisions with reduced computational complexity, enabling real-time implementation.

\section{Performance Evaluation}
\label{Performance Evaluation}

\begin{table}[t]
\centering
\caption{System and Simulation Parameters}
\small
\setlength{\tabcolsep}{3pt}
\renewcommand{\arraystretch}{1.0}
\begin{tabular}{|p{2.2cm}|p{1.6cm}|p{2.2cm}|p{1.6cm}|}
\hline
\textbf{Parameter} & \textbf{Value} & \textbf{Parameter} & \textbf{Value} \\
\hline
User speed & 0--1.5 m/s & GM memory parameter & 0.8 \\
\hline
Velocity variance & 0.1 & Direction variance & 0.05 \\
\hline
Orientation variance & 0.02 & Vertical separation & 3 m \\
\hline
$w$ & 5 $\mu$m & Wavelength & 1550 nm \\
\hline
RIN PSD & $-155$ dB/Hz & PD responsivity & 0.7 A/W \\
\hline
Noise figure & 5 dB & Bandwidth & 1.5 GHz \\
\hline
\multicolumn{4}{|c|}{\textbf{LSTM Parameters}} \\
\hline
Sequence length & 5 steps & Hidden units & 64 \\
\hline
Layers & 2 & Learning rate & $10^{-3}$ \\
\hline
Batch size & 64 & Epochs & 100 \\
\hline
Dropout rate & 0.2 & & \\
\hline
\end{tabular}
\label{tab:combined_parameters}
\end{table}

This section evaluates the performance of the proposed MAPC framework through numerical simulations. The objective is to quantify the benefits of incorporating mobility prediction into transmit power allocation in terms of prediction accuracy, energy efficiency, and robustness to user mobility. MAPC is compared against three benchmark strategies. First, Conventional Power Control (CPC), which determines transmit power using only instantaneous channel state information without accounting for future mobility. Second, Conservative Power Control (ConsPC), which allocates power based on the current channel estimate with an added fixed safety margin to compensate for potential degradation. Third, Reactive Power Control (RPC), which adjusts transmit power only after link degradation or quality-of-service violations are observed. %These schemes differ in how channel information is utilised: MAPC proactively exploits predicted future channel states, CPC relies solely on current conditions, ConsPC accounts for uncertainty through fixed safety margins, and RPC responds only after performance degradation. 

An indoor OWC network is deployed in a \(5 \times 5 \times 3\) m\(^3\) room with \(A=12\) APs and a varying number of mobile users. Only line of sight links are considered, as it is assumed that VCSEL beams concentrate optical power along the direct path, making diffuse components negligible. The network operates with a control interval of \(100\) ms, during which the CU performs mobility prediction, channel estimation, and transmit power optimisation. User mobility follows a GM process, with walking speeds ranging from 0 to 1.5 m/s, while device orientation varies dynamically to emulate realistic handheld usage. In addition, behaviour-driven variations such as abrupt changes in direction, speed fluctuations, and device tilting are introduced to capture realistic user dynamics. Results are averaged over multiple independent mobility simulations. 

\begin{figure}[t]
\centering 
\includegraphics[width=1\columnwidth]{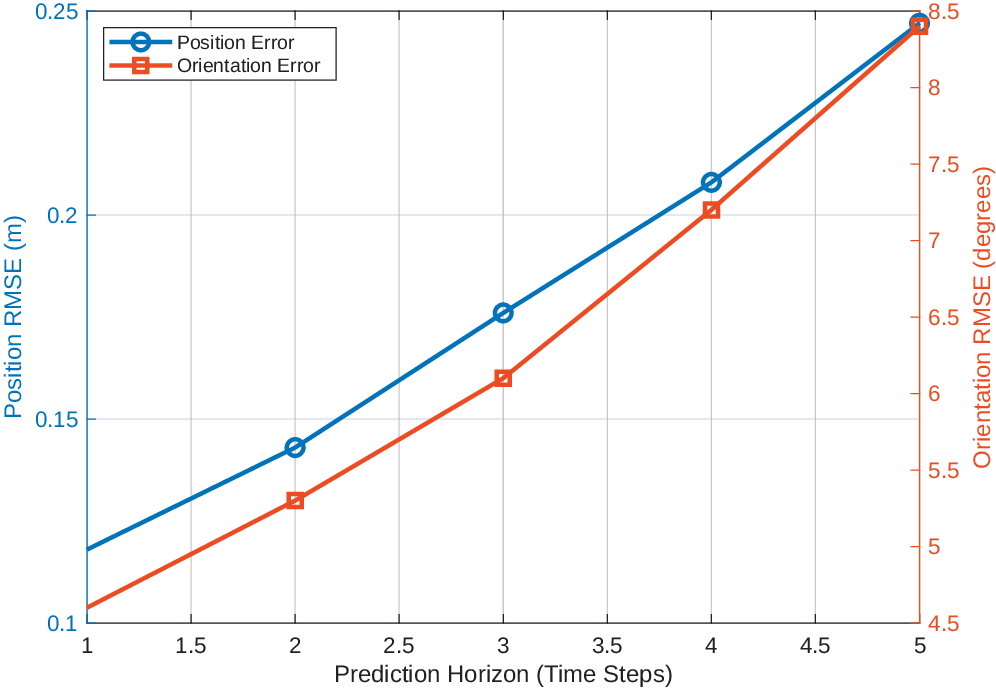}
\caption{Mobility Prediction Accuracy.} 
\label{fig:Prediction Accuracy} 
\end{figure}

Mobility prediction accuracy is evaluated using the root mean squared error (RMSE) between the predicted and actual mobility states. Fig.~\ref{fig:Prediction Accuracy} shows the prediction error for user position and device orientation versus the prediction horizon, defined as the number of future time steps for which mobility is predicted. As shown, the prediction error increases gradually with the prediction horizon. However, short-term predictions remain sufficiently accurate to estimate future channel conditions, enabling reliable power allocation.

\begin{figure}[t]
\centering 
\includegraphics[width=0.9\columnwidth]{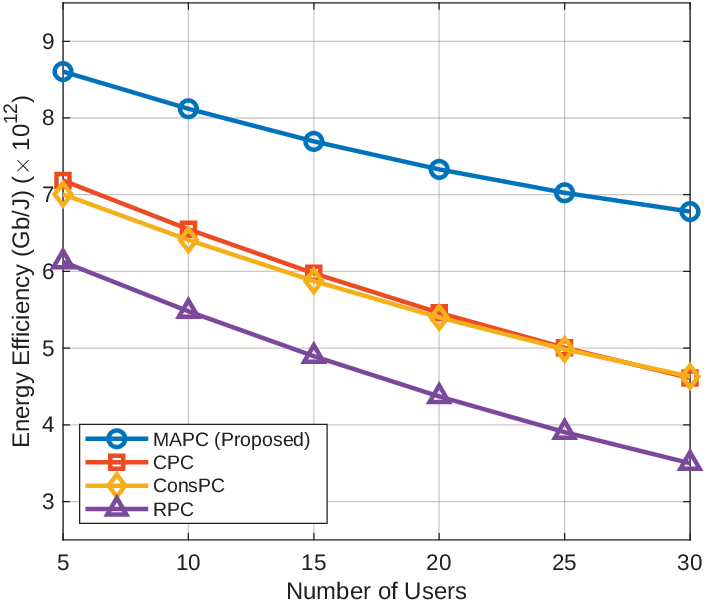}
\caption{Energy Efficiency Performance at user speed of 1m/s.} 
\label{fig:Energy Efficiency Performance} 
\end{figure}

Fig.~\ref{fig:Energy Efficiency Performance} shows the EE performance of MAPC against the benchmark schemes for users at a fixed speed of 1 m/s. The results demonstrate that MAPC consistently achieves higher EE than the benchmark schemes. By considering predicted mobility states, the CU anticipates future channel conditions and allocates only the power required to satisfy expected user demand. In contrast, CPC relies solely on instantaneous channel observations, ConsPC introduces a conservative safety margin that increases power consumption, and RPC reacts to channel degradation only after it occurs.

\begin{figure}[t]
\centering 
\includegraphics[width=0.9\columnwidth]{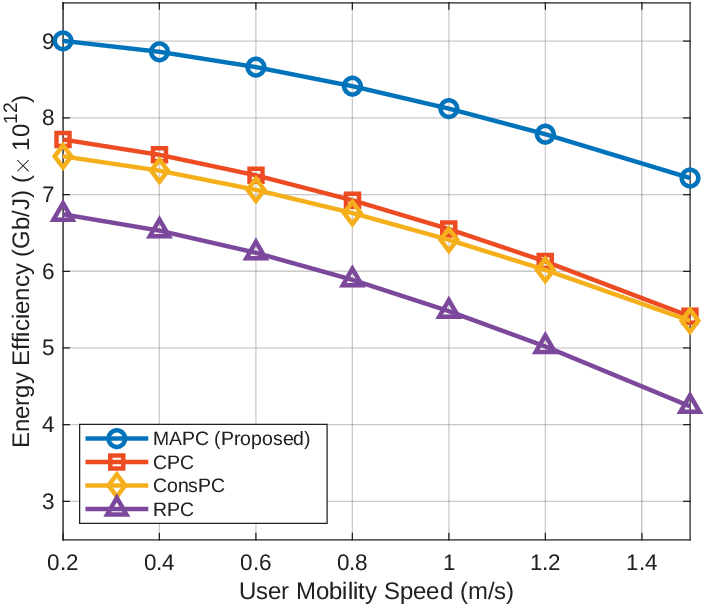}
\caption{Impact of Mobility Speed for 10 users.} 
\label{fig:Impact of Mobility Speed} 
\end{figure}

To evaluate robustness under dynamic user movement, the average walking speed of 10 users is varied between \(0.2\) m/s and \(1.5\) m/s. Fig.~\ref{fig:Impact of Mobility Speed} shows the resulting variation in EE. Although EE decreases for all schemes as mobility speed increases due to more frequent beam misalignment and stronger channel fluctuations, the proposed framework exhibits significantly smaller degradation. This improvement is achieved because the CU anticipates mobility-induced channel variations and adjusts transmit power proactively.

\section{Conclusions}
\label{Conclusion and Discussion}
This paper presented a mobility-aware power control framework for VCSEL-based indoor OWC networks. The proposed approach integrates mobility prediction, and channel estimation to proactively allocate transmit power according to anticipated network conditions. User mobility is modelled using a hybrid GM-ANN framework that captures both motion dynamics and device orientation changes, enabling prediction of future optical channel conditions. Simulation results show that incorporating mobility-aware channel prediction improves network EE compared with conventional mobility-agnostic power allocation schemes while maintaining robust performance under increasing user mobility. MAPC consistently outperforms the benchmark schemes, achieving average EE gains of approximately 30–35 \%  over CPC and ConsPC, and 45–65\% over RPC across different user loads and mobility conditions. These results highlight the importance of integrating mobility prediction into resource allocation for OWC systems. Future work will focus on evaluating the proposed framework under more realistic indoor environments, including the effects of user blockage, shadowing from moving objects, and multipath reflections from walls and furniture. In addition, future work will explore joint mobility and traffic demand prediction for adaptive resource allocation across different user services.

\bibliographystyle{IEEEtran}
\bibliography{References}

\end{document}